\documentclass{aa}

\usepackage[varg]{txfonts}
\usepackage{graphicx}
\usepackage[colorlinks=true,citecolor=blue]{hyperref}
\usepackage{enumitem}

\begin{document}

\title{Cluster counts: Calibration issue or new physics?}
\titlerunning{Cluster counts: Calibration issue or new physics?}
\author{Ziad Sakr\inst{1,2}
\and Stéphane Ili\'c\inst{3,1}
\and Alain Blanchard\inst{1}
\and Jamal Bittar\inst{2}
\and Wehbeh Farah\inst{2}}
\authorrunning{Sakr et al.}
\institute{IRAP, Université de Toulouse, CNRS, CNES, UPS, Toulouse, France
\and
Universit\'e St Joseph; UR EGFEM, Faculty of Sciences, Beirut, Lebanon
\and
CEICO, Institute of Physics of the Czech Academy of Sciences, Na Slovance 2, Praha 8 Czech Republic \\[0.2cm]
\email{\scriptsize zsakr@irap.omp.eu, ilic@fzu.cz, alain.blanchard@irap.omp.eu, principal@tes.edu.lb, wehbeh.farah@usj.edu.lb}
}
\date{Received XXXX-XX-XX; accepted XXXX-XX-XX}

\abstract{
In recent years, the amplitude of matter fluctuations inferred from low-redshift probes has been found to be generally lower than the value derived from cosmic microwave background (CMB) observations in the $\Lambda$CDM  model. This tension has been exemplified by Sunyaev-Zel'dovich and X-ray cluster counts which, when using their Planck standard cluster mass calibration, yield a value of $\sigma_8$ , appreciably lower than estimations based on the latest Planck CMB measurements.

In this work we examine whether non-minimal neutrino masses can alleviate this tension substantially. We used the cluster X-ray temperature distribution function derived from a flux-limited sample of local X-ray clusters, combined with Planck CMB measurements. These datasets were compared to $\Lambda$CDM predictions based on recent mass function, adapted to account for the effects of massive neutrinos. Treating the clusters mass calibration as a free parameter, we examined whether the data favours neutrino masses appreciably higher than the minimal 0.06~eV value. Using Markov chain Monte Carlo methods, we found no significant correlation between the mass calibration of clusters and the sum of neutrino masses, meaning that massive neutrinos do not noticeably alleviate the above-mentioned Planck CMB--clusters tension. The addition of other datasets (BAO and Ly-$\alpha$) reinforces those conclusions. 

As an alternative possible solution to the tension, we introduced a simple, phenomenological modification of gravity by letting the growth index $\gamma$ vary as an additional free parameter. We find that the cluster mass calibration is robustly correlated with the $\gamma$ parameter, insensitively to the presence of massive neutrinos or/and additional data used. We conclude that the standard Planck mass calibration of clusters, if consolidated, would represent  evidence for new physics beyond $\Lambda$CDM with massive neutrinos.
}


\maketitle


\section{Introduction}\label{sec:1-Intro}   

The accumulation of high-quality data over the last three decades allows us now to consider cosmology as a precision science \citep{1992ApJ...396L...1S, 1998AJ....116.1009R,   1999ApJ...517..565P, 2005ApJ...633..560E, 2013ApJS..208...20B, 2014A&A...568A..22B, 2017MNRAS.470.2617A, 2018arXiv180706209P}. The standard model of cosmology, the $\Lambda$CDM paradigm, successfully reproduces the vast majority of those observations \citep{2004PhRvD..69j3501T, 2008ApJ...686..749K, 2010A&ARv..18..595B} and the values of its associated cosmological parameters are now very well constrained \citep{2016A&A...594A..13P}. The $\Lambda$CDM model succeeds not only in explaining the observed properties of our present Universe (supernovae, baryon acoustic oscillations) and its early stages (cosmic microwave background fluctuations) beyond the standard Big Bang picture (expansion, CMB spectrum, Big Bang nucleosynthesis) but also in predicting some of these specific properties \citep{2003A&A...412...35B}. However, recent results revealed some tensions between this standard theory and observables of the late Universe. One particular result has attracted a lot of attention with the first data release of the Planck satellite: the measurement of the abundance of galaxy clusters detected through their imprint on the CMB by the Sunyaev-Zel'dovich (SZ) effect \citep{2014A&A...571A..20P}. Taken at face value, the observed SZ-cluster number counts -- using a specific calibration of cluster masses -- are significantly lower than predicted by the $\Lambda$CDM model when using the cosmological parameters derived from CMB data. This leads to an appreciable difference in the derived value of the $\sigma_8$ parameter, which characterizes the current amplitude of matter fluctuations. Several local probes also lead to some similar tensions: measurements of the linear growth rate of structures through weak lensing \citep{2013MNRAS.432.2433H, 2017MNRAS.471.4412K, 2016PhRvD..94b2001A} and redshift-space distortions \citep{2017MNRAS.469.1369S} that appear to be consistently lower than the Planck-normalized $\Lambda$CDM predicted values. However, those apparent discrepancies have not yet reached a ``worrying'' level of significance, and may also be potentially affected by systematic effects \citep{2007MNRAS.376...13M, 2014MNRAS.439...48A}. It should also be noted that the comparison of the most recent results from the large-scale Dark Energy Survey \citep{PhysRevD.98.043526} and the latest Planck CMB observations \citep{2018arXiv180706209P} show a less severe tension.

The origin of the discrepancy between CMB cosmology and cluster counts measurements remains an open question. Considering the current observations by the Planck satellite (as well as the Atacama Cosmology Telescope and South Pole Telescope on the ground), it is fairly reasonable to assume that most of the tension should not originate from CMB observations. Potential biases or systematic effects might remains, but it appears unlikely given the exquisite quality of the current data and the meticulous care taken in their analysis (see \citealt{2016A&A...594A...1P, 2014A&A...571A..16P}, as well as further analyses by \citealt{2017A&A...602A..41C, 2017A&A...597A.126C} and consistency checks with previous data from the Wilkinson Microwave Anisotropy Probe by \citealt{2017A&A...607A..95P}).

However, the situation is less clear for clusters and their cosmological analysis. On the one hand, a robust theoretical framework has been available for predicting cluster counts since the seminal work of \citet{1974ApJ...187..425P}, allowing us to compute the so-called halo mass function -- the abundance of dark matter halos as a function of their mass. Since then, this framework has been consolidated thanks to the use of N-body simulations \citep{1999MNRAS.308..119S,2001MNRAS.321..372J,2008ApJ...688..709T,2011MNRAS.410.1911C,2013MNRAS.433.1230W,2016MNRAS.456.2486D}. On the other hand, a variety of issues can in practice plague the analysis of clusters samples, such as unaccounted or incorrectly accounted statistical biases (e.g. Eddington and Malmquist biases), or improper selection functions \citep[see e.g.][]{2016JCAP...08..013B,2017MNRAS.464.2270A}. However, one major obstacle in particular stands out: the theoretical mass function predicts cluster abundances as a function of their total mass, but the latter is not a directly observable quantity. Consequently, proxies are required and are related more or less directly to the cluster mass: luminosity, X-ray temperature, and weak lensing for example. These observables are then related to the mass through so-called scaling laws or relations that need to be carefully calibrated. A certain number of assumptions are often required, and a consensus has yet to be reached in many cases for such relations. A crucial point is that differences in the scaling laws and their normalization can in turn lead to differences in the cosmological results that are obtained from a given dataset. A bias in the scaling laws can thus propagate into biases in the inferred parameters \citep{2005A&A...436..411B}.

If the tension between early and late observables were confirmed, we might have to consider as a consequence extensions or alternatives to the standard $\Lambda$CDM model of cosmology. Since the CMB itself is mostly sensitive to the physics of the early Universe, one can reconcile it with clusters observations by introducing a modification that only has a significant impact at late times. More specifically, a new theory with a lower growth rate of structures would predict a lower abundance of clusters, in better agreement with the data. Modifications of the growth rate can result from different physical origins: one possibility is to add mass to neutrinos in the standard cosmology (instead of approximating them as massless). Among other effects, massive neutrinos indeed slow down the growth of matter perturbations during the matter- and dark-energy-dominated era on scales smaller than their free-streaming length (see e.g. \citealt{2012arXiv1212.6154L} for a review on neutrinos in cosmology).

In the present paper, we first examine whether the aforementioned discrepancy between clusters and CMB cosmology can be solved by introducing massive neutrinos. Combining the temperature distribution function from a flux-limited sample of local X-ray clusters with the latest CMB measurements from the Planck satellite, we performed a Bayesian analysis through Monte Carlo Markov Chains (MCMC), using not only the parameters of the standard model but also the neutrino masses and the cluster mass calibration as free parameters. We later introduced a phenomenological modification of the growth rate of structures, modelled by the so-called growth index $\gamma$ \citep[see e.g.][]{1998ApJ...508..483W,2005PhRvD..72d3529L} as an additional degree of freedom. We tested the robustness of our conclusions when including additional constraints and datasets, namely baryon acoustic oscillations (BAO) and Lyman-alpha (Ly-$\alpha$) forests, which probe the late Universe at redshifts higher than our cluster sample.

In Sect.~\ref{sec:2-AbunCosmo}, we describe the formalism used for predicting clusters abundances, as well as the extensions to the standard model we examine and their impact on cluster counts. In Sect.~\ref{sec:3-DataMeth}, we detail the datasets used in this work, and the implementation of the MCMC analysis to sample the posterior probability distribution function. We present and discuss our results in Sect.~\ref{sec:4-Res} and summarize our conclusions in Sect.~\ref{sec:Conclusions}.


\section{Cluster abundance and cosmology}\label{sec:2-AbunCosmo}


\subsection{The halo mass function}\label{ssec:2.1-HMF}

In the standard theory of structure formation, primordial small inhomogeneities grew by gravitational instability in an expanding universe \citep{1933ASSB...53...51L}. At early times, the amplitude of fluctuations is small and their growth can be described by linear theory. Clusters are, however, non-linear objects in the sense that their contrast density is much larger than one. The formation of these objects therefore cannot be tracked directly by linear theory. However, they are believed to result from collapsing regions in the gravitational instability picture as derived by Jeans in static Newtonian theory \citep{1902RSPTA.199....1J}. The dynamics of spherical regions in general relativity was provided by \citet{1933ASSB...53...51L}. The non-linear spherical model allows us to link the collapse of non-linear objects to the sole condition that their linear amplitude is larger that some threshold $\delta_c$.

Derivation of the mass function of cosmological structures from initially Gaussian fluctuations was first addressed by \cite{1974ApJ...187..425P}. Under general hypotheses of self-similarity, the exact mass function can be written in a simple form \citep{1992A&A...264..365B},
\begin{equation}\label{eq:nm}
  n(m) = -\frac{\rho_0}{m^2}\frac{\mathrm{d} \ln \nu}{\mathrm{d} \ln m}\nu \mathcal{ F}(\nu) 
,\end{equation}
where $\rho_0$ is the mean matter density today, and $\nu = \delta_{c}(z)/\sigma(m)$ is the normalized amplitude of fluctuations. Within a sphere of comoving radius $R$ that contains mass $m=4\pi\rho_0 R^3/3$, $\sigma^2(m)$ is the variance of the linear density perturbations:
\begin{equation}\label{eq:s2m}
  \sigma^2(m)=\sigma^2(R)=\frac{1}{2\pi^2}\int_0^\infty k^2 P(k) W^2(kR) dk
,\end{equation}
where $P(k)$ is the linear power spectrum and $W(kR)$ is the Fourier transform of the top-hat window function. As mentioned before, $\delta_c$ represents the critical value of the initial overdensity that is required for collapse at $z$, computed using the spherical collapse model. In the most general case, this quantity is redshift-dependent although weakly in the $\Lambda$CDM paradigm \citep{1996MNRAS.280..638K} and with a weak dependency on cosmological parameters. In the following, we use the fitting formula of \citet{1996MNRAS.280..638K}.

The original form of the function $\mathcal{F}$ derived by Press \& Schechter is
\begin{equation}
   \mathcal{F}_{\rm PS}(\nu) =\sqrt{\frac{2}{\pi}}\ \ \exp\left[-\frac{\nu^2}{2}\right].
\end{equation}
A more refined determination of $\mathcal{F}$ has been the subject of numerous investigations such as \citet[][ST99 hereafter]{1999MNRAS.308..119S} who investigated the consequences of the non-sphericity of the collapse, while \cite{1991ApJ...379..440B} used the peak formalism and derived the so-called ``excursion set" theory as a generalization of the Press \& Schechter formalism. Afterwards, continuous improvements of N-body simulations have allowed a more accurate determination for the mass function (cf. citations throughout this work). Two different approaches can be found in the literature:
\begin{itemize}[noitemsep,topsep=0pt]
  \item[i)] the first approach uses the ST99 formula (or small variations thereof), which is based on the historically-first functional form of Press \& Schechter,
  \begin{equation}
    \nu \mathcal{ F}_{\rm ST}(\nu) =A\sqrt{\frac{2a}{\pi}}\left[1+\left(\frac{1}{a\nu^2}\right)^p\right]\ \exp\left[-\frac{a\nu^2}{2}\right],
  \end{equation}
  where $A$, $a,$ and $p$ were parameters originally fitted by ST99 on an N-body simulation with Einstein-de Sitter cosmology;
  \item[ii)] the second approach formulates the mass function in terms of $\sigma$ rather than $\nu$ \citep[see e.g.][]{2001MNRAS.321..372J,2006ApJ...646..881W}. A widely used form is that of \citet[][T08 hereafter]{2008ApJ...688..709T},
  \begin{equation}
    \nu \mathcal{ F} = f_{\rm T}(\sigma) =A\left(\left(\frac{b}{\sigma}\right)^a + 1\right) \exp\left[-\frac{c}{\sigma^2}\right],
  \end{equation}
  where $A$, $a$, $b,$ and $c$ were fitted on $\Lambda$CDM N-body simulations.
\end{itemize}

This second formulation breaks the self-similar nature of the mass function, essentially meaning that the threshold $\delta_c$ dependence on cosmology as well as the virial reference are irrelevant. This might well be because both the concordance cosmological model and its power spectrum are not self-similar. Indeed, several authors recently claimed a break of the self-similarity of the mass function (\citealt{2008ApJ...688..709T}, see also \citealt{2011MNRAS.410.1911C}).

However, using very high-resolution simulations, \citet[][hereafter D16]{2016MNRAS.456.2486D} recently found that the scaling of the mass function implied by Eq.~\ref{eq:nm} still holds, provided that one uses the spherical overdensity algorithm and the virial mass density contrast for halo definition. Furthermore, they showed that the ST99 formula provides an accurate fit at all relevant redshifts and a wide variety of $\Lambda$CDM cosmologies, after revising the values of the free parameters of the fitting function. The authors provide as well a second set of values to be used specifically for an optimal fit on clusters scales. In the present work, we checked that these two sets lead to virtually identical results when applied to our data. In this work we use the D16 formula of the mass function,
\begin{equation}
   \mathcal{ F}_{\rm D}(\nu) =A\sqrt{\frac{2a}{\pi}}\left[1+\left(\frac{1}{a\nu^2}\right)^p\right]\  \ \exp\left[-\frac{a\nu^2}{2}\right]
,\end{equation}
with $A=0.3295 $, $a=0.7689,$ and $p=0.2536$ with a redshift-dependent function for $\delta_c$ appropriate for $\Lambda$CDM models.

In order to compare the theoretical halo mass function to the actual measured abundance of galaxy clusters, we need a relation between the cluster mass $m$ entering the mass function, and the clusters observable $O$ considered (e.g. SZ signal, temperature). Furthermore, some dispersion is expected in the relation, which can be taken into account by writing
\begin{equation}\label{eq:scallaw0}
  N(>O) = \int_0^{+\infty}p(>\!O|m) \ n(m) \ dm
,\end{equation}
where $p(>\!\!O|m)$ represents the probability that a cluster of mass $m$ will be observed with a value of the observable greater than $O$. A convenient way to take the dispersion into account is to assume a log-normal probability distribution for $O$, which leads to a (positive) offset in the calibration of mass-observable relation \citep{2000A&A...362..809B}. In the present study we use the cluster temperature as observable and rely on the following relation between temperature and mass, assuming a standard power-law scaling relation:
\begin{equation}\label{eq:scallaw}
  T=A_{T-M}(h\, M_\Delta)^{2/3}\left(\frac{\Omega_{m} \Delta(z)}{178}\right)^{1/3}(1+z)
,\end{equation}
where $A_{T-M}$ is the normalization parameter, $\Delta$ is the density contrast chosen for the definition of a cluster, expressed with respect to the total background matter density\footnote{Equation~\ref{eq:scallaw} is different when the contrast density $\Delta_c$ is expressed with respect to the critical density at redshift $z$.} of the Universe at redshift $z$, and $M_\Delta$ is the mass of the cluster according to the same definition. We note that the $2/3$ exponent is consistent with the existing data \citep{2015A&A...582A..79I}. The dispersion is taken into account in the calibration according to the earlier remark; more details on this procedure can be found in \citet{2015A&A...582A..79I}. Relation (\ref{eq:scallaw0}) can then be used to determine the integrated temperature function and becomes:
\begin{equation}
  N(>T) =  \int_{M(T)}^{+\infty} n(m) dm 
.\end{equation}
The calibration of the relation $A_{T-M}$ is a subject of strong debate: standard mass estimates are based on hydrostatic assumptions although these are subject to theoretical uncertainties  \citep{1997ApJ...487...33B}. Calibration of X-ray telescopes, in particular between XMM and Chandra, is also an issue although not regarded as large enough possibly to solve the discrepancy \citep{2015A&A...575A..30S,2015MNRAS.448..814I}. Furthermore, hydrodynamical simulations have shown that gas in clusters is not in hydrostatic equilibrium \citep{1998ApJ...495...80B, 2007ApJ...655...98N, 2008A&A...491...71P, 2010A&A...514A..93M}. This has lead to the introduction of an encompassing ``mass bias'' (referred to as $1-b$ in the literature or more properly as the $B= 1/(1-b)$ parameter, cf. \citealt{2018MNRAS.480.3928M}) defined as the ratio of the mass proxy used to establish the scaling relation and the true mass \citep[see e.g.][]{2014A&A...571A..29P}. 

Several attempts have been made to determine $A_{T-M}$ both from theoretical considerations and simulations, with inferred values varying from $\sim3$ to $\sim6$ \citep{2005A&A...436..411B}. Given the above tension, obtaining a reliable mass proxy with a well-determined calibration has become a critical issue in clusters studies and has been the focus of many works in the recent literature \citep{2016MNRAS.456.2301W,2018arXiv180706209P}. \citet{2015A&A...582A..79I} used a different approach in the context of the $\Lambda$CDM model, treating $A_{T-M}$ as an additional free parameter to be determined, and constrained it using MCMC techniques with a robust local X-ray clusters sample and Planck CMB measurements as data. In the present work, we follow the same approach and add $A_{T-M}$ as a free parameter in the analysis, in a more general cosmological context.



\subsection{Neutrinos and the growth of structures}\label{ssec:2.2-nu}

In the present work, we investigate ways to alleviate the tension between early and late cosmological probes, exemplified by cluster abundance measurements. As mentioned earlier, lowering the growth rate of structures in the Universe can reconcile the two datasets. Massive neutrinos -- beyond the ``minimal mass'' attributed to them in the current standard cosmological model -- offer a possible solution to the tension. Indeed, their presence can alter the aforementioned growth rate, damping the amplitude of matter fluctuations on scales smaller than their free-streaming length. This is further motivated by the fact that neutrinos are experimentally known to be massive, with at least two species being non-relativistic today.

We briefly recall here some elements of neutrino physics and their influence in cosmology, considering three families with non-zero, degenerate masses, which will be our assumption throughout the rest of this work \citep[see][for a complete review]{2012arXiv1212.6154L}. After decoupling from the rest of the matter-energy content in the early universe ($\sim$ 1s after the Big Bang), massive neutrinos remain relativistic for an extended period of time, and as such are part of the radiation content of the Universe. Their energy density can be expressed as a function of the photon energy density and the effective number of neutrinos $N_{\rm eff}$ ($\simeq$3.046 in our case), defined as
\begin{equation} 
\label{eq:radnu}
  \rho_\nu = N_{\rm eff} \frac{7}{8} \left(\frac{4}{11}\right )^{4/3} \rho_\gamma.
\end{equation}
This equation is valid when neutrino decoupling is complete and holds as long as all neutrinos are relativistic. At later times, massive neutrinos become non-relativistic and can be considered as part of the matter content of the Universe. Their energy density today in units of the critical density can be approximated as
\begin{equation}
  \Omega_\nu = \frac{\rho_\nu}{\rho_c^0} = \left(\frac{N_{\rm eff}}{3}\right)^{3/4} \frac{\sum m_\nu}{94.07 \, h^2 \, {\rm eV}}
,\end{equation}
where $\sum m_\nu$ is the sum of all three neutrino masses.

The presence of neutrinos (massive or not) affects cosmological observables in several ways. As part of the total energy content of the Universe, they influence its background evolution to a varying degree, depending on their properties (number, masses, and so on). As a consequence all observables based on distance measurements will be affected, such as supernovae, baryon acoustic oscillations, and the angular scale of the sound horizon at last scattering (as measured by the position of CMB peaks). Beyond their effect on the locations of the peaks, neutrinos also have an effect around the first acoustic peak of the CMB, which is due to the early integrated Sachs-Wolfe (ISW) effect. Moreover, the latest CMB data from the Planck satellite now allow us to probe another signature of neutrinos in the CMB, namely their effect through gravitational lensing, which dampens the amplitude of the acoustic peaks.

More generally, as mentioned before, massive neutrinos affect the growth of structure, which in turn affects many observables: gravitational lensing, galaxy clustering, as well as the abundance of galaxy clusters. This can be expressed through their effect on the so-called matter power spectrum,
\begin{equation}
  \label{eq:defPk}
  P_{\rm m}(k,z)=<|\delta_{\rm m}(k,z)|^2>
,\end{equation}
where $\delta_{\rm m} = \delta\rho_{\rm m}/\bar{\rho}_{\rm m}$ represents the total matter overdensity. In practice, the matter power spectrum at any redshift can be written as
\begin{equation} 
  \label{eq:pk}
  P_{\rm m}(k,z) =  P_{\rm prim}(k) T^2_{\rm m}(k,z) 
,\end{equation}
where $P_{\rm prim}(k)$ is the near scale-invariant primordial power spectrum, and $T_{\rm m}$ is a so-called ``transfer function'' that encapsulates all the details of the growth of structures. This function can be split into different contributions corresponding to the various forms of matter:
\begin{equation}
   T_{\rm m}= \frac{\Omega_{\rm cdm} T_{\rm cdm} + \Omega_{\rm b} T_{\rm b} + \Omega_{\nu} T_{\nu}}{\Omega_{\rm cdm}+\Omega_{\rm b}+\Omega_{\nu}}
\end{equation}
where the ${\rm cdm}$, ${\rm b,}$ and ${\nu}$ subscripts respectively refer to cold dark matter, baryons, and neutrinos. Then, the influence of neutrinos depends on the wavenumber $k$ considered:
\begin{itemize}[noitemsep,topsep=0pt]
  \item on scales larger than a certain threshold (roughly proportional to the inverse square root of their mass), neutrino free-streaming can be ignored and neutrino perturbations are indistinguishable from CDM perturbations. On those scales, the matter power spectrum $P_{\rm m}(k,z)$ can be shown to depend only on the matter density fraction today (including neutrinos) and the primordial perturbation
  spectrum.
  \item on scales smaller than the free-streaming length, massive neutrinos do not cluster, that is, $\delta_{\nu} \ll \delta_{\rm cdm}$ ($\sim \delta_{\rm b}$). Consequently, even if the evolution of $\delta_{\rm cdm}$ was not affected by neutrinos, the power spectrum would be reduced by a factor $(1-f_\nu)^2$ where $f_\nu\equiv\Omega_\nu/\Omega_{\rm m}$. In practice, the growth rate of $\delta_{\rm cdm}$ is reduced through an absence of gravitational back-reaction effects from free-streaming neutrinos. At low redshift, the matter power spectrum is thus affected by a step-like suppression that starts around the free-streaming scale and saturates at higher wavenumbers ($k\sim 1$ h/Mpc) with a constant amplitude $\Delta P_{\rm m}(k)/P_{\rm m}(k)\simeq-8f_\nu$.
\end{itemize}

Although we focused here on the effects of a non-zero mass, the growth of structure is also sensitive to the effective number of neutrino species. Stringent experimental limits have determined the number of so-called ``active'' neutrinos (sensitive to weak interactions) to be equal to three, but leave room for additional, ``sterile" neutrinos species that interact only through gravity and can leave an imprint on cosmological observables \citep[see][for a detailed discussion] {2013neco.book.....L}. In the present work, we report our results on such scenarios in Appendix~\ref{sec:Appendix-A}.


\subsection{Effects of modifications of gravity}\label{ssec:2.3-MG}

An alternative approach to reconcile early and late probes is to consider modifications to the standard theory of gravity. Indeed, such modifications can potentially lower the growth rate of structures by effectively reducing the strength of the gravitational force. In this context, one can either introduce a new full theory of gravitation such as $f(R)$ or Galileons \citep[see e.g.][for a comprehensive review]{2012PhR...513....1C} or apply a phenomenological modification to the equations governing structure formation. In the present work, we follow the latter approach by modifying directly the growth rate of structure.

Let us recall here a few elements of linear perturbation theory. In the $\Lambda$CDM paradigm, after radiation-matter equality, the following second-order equation governs the growth of the matter perturbations $\delta_{\rm m}$:
\begin{equation}
  \delta_{\rm m}'' + [2+(\ln H)'] \delta_{\rm m}' = \frac{3}{2} \Omega_{\rm m}(a) \delta_{\rm m},
\end{equation}
where primes denote derivatives with respect to $\ln a$ and $\Omega_{\rm m}(a) \equiv 8 \pi G \rho_{\rm m}(a)/3 H^2$. We usually define the so-called growth factor $D(a)$ with
\begin{equation}
  \delta_{\rm m} (a) = \delta_{\rm m,0} D(a),
\end{equation}
where $\delta_{\rm m,0}$ is often taken by convention to be the matter density contrast today at $a=1$, thus $D(a)=1$ today. The growth factor in $\Lambda$CDM is very well approximated by the expression \citep{1980lssu.book.....P} 
\begin{equation}
  D(a) \approx \exp \left(\int_1^a \frac{da'}{a'} \Omega_{\rm m}(a')^\gamma \right)
\end{equation}
or equivalently, that the so-called growth rate $f$ defined as
\begin{equation}
  f \equiv \frac{d \ln D}{d \ln a} 
\end{equation}
is well approximated by
\begin{equation}
  \label{eq:fgam}
  f \approx \Omega_{\rm m}(a)^\gamma 
,\end{equation}
where $\gamma$ is called ``growth index'' and found to be $\sim 0.545$ in $\Lambda$CDM. A simple and direct way to introduce a phenomenological modification to the growth rate is to define it exactly as in Eq.~\ref{eq:fgam} and consider $\gamma$ as an additional free parameter than can differ from its $\Lambda$CDM value \citep[an idea first introduced by][]{2005PhRvD..72d3529L}. This approach has been adopted many times in the literature, both when analysing current data as well as in the context of forecasts for future missions \citep[cf.][for the future Euclid satellite]{2011arXiv1110.3193L}. However, using only a single number to parametrize deviations of the growth rate not only confines us to a limited number of scenarios, but the remaining ones also struggle to reproduce the behaviour of any actual, realistic theory of modified gravity. As a consequence, a number of authors explored instead the use of $\gamma$ as a varying function of time by expressing it as a Taylor expansion (with respect to $z$, $a$, $\ln a$, ...) truncated to some order \citep{2008PhRvD..78l3010G}. Then, the coefficients of the expansion $\gamma_0$, $\gamma_1$, and so on become additional free parameters in the analysis. Some authors have derived recipes to link such sets of coefficients with actual full theories of modified gravity \citep{2014JCAP...05..042S}. In the present work we limit our analysis to the original $\gamma$ parameter; further parametrizations will be explored in a future work.


\subsection{Impact on cluster observables}\label{ssec:2.4-ClusObs}

\begin{figure}[t]
  \centering
  \includegraphics[width=\columnwidth]{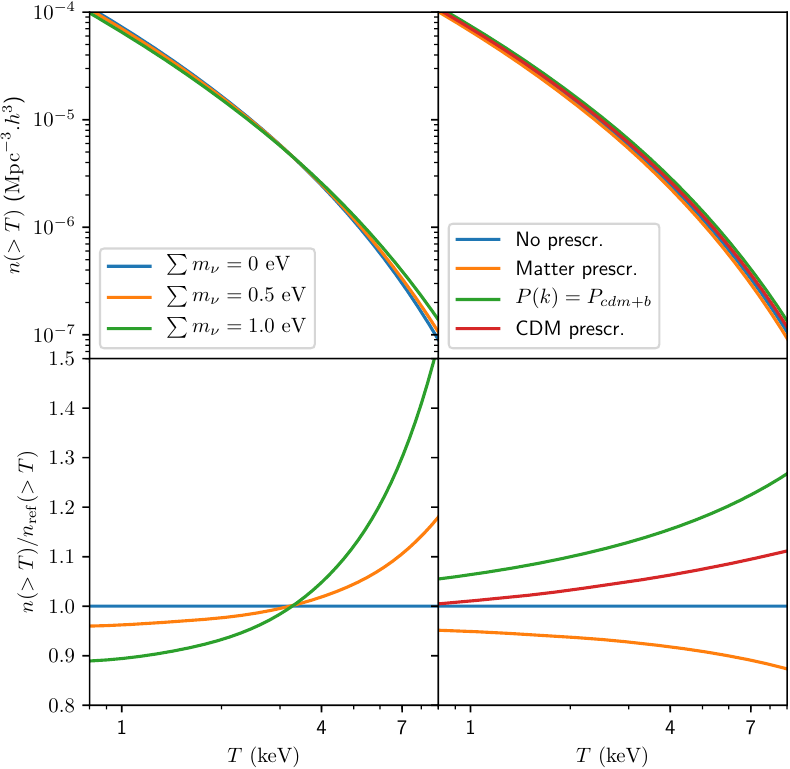}
  \caption{Temperature distribution functions (top panels) and ratios with respect to the blue reference model (bottom panels). The left panels illustrate the effect of massive neutrinos (no neutrino prescription used, $\sigma_8$ and $\Omega_m$ kept constant) while the ones on the right show the effects of the various steps of the neutrino prescription for a given neutrino mass ($\sum m_\nu=0.5$ eV).}
  \label{fig:dndT}
\end{figure}

To conclude this section, we illustrate the effects of massive neutrinos and a modification of gravity on our observable of interest: the abundance of local clusters as a function of mass. Both phenomena are related to cluster counts mainly through their influence on the linear growth rate of structure, which in turn affects the matter power spectrum $P_{\rm m}(k,z)$ and the variance of the associated matter density fluctuations defined in Eq.~\ref{eq:s2m}. 
The two left panels of Fig.~\ref{fig:dndT} show the impact of different values of the neutrino masses -- with all other cosmological parameters fixed \footnote{More precisely, amongst the standard parameters used in CMB studies, $\tau$, $\theta_{MC}$ , and $n_s$ are kept fixed, while $\Omega_bh^2$, $\Omega_ch^2$ , and $A_s$ are adjusted to give a constant $\Omega_m$ and $\sigma_8$ for all neutrino masses considered.} -- on the cumulative temperature mass distribution. With the $\sigma_8$ parameter fixed, the three neutrino masses naturally yield the same cluster abundance at the temperature corresponding to the 8 Mpc$/h$ scale, here $\sim$ 4 keV.

We mention here an important point regarding neutrinos and cluster counts. Recent works in the literature \citep{2013JCAP...12..012C,2014JCAP...02..049C} based on high-resolution N-body simulations showed that theoretical cluster counts computed the ``traditional" way are a poor fit to numerical simulations that include massive neutrinos. The same authors provide a new prescription for the theoretical number counts in the presence of massive neutrinos that significantly improves the fit and consists in the two following steps:
\begin{itemize}[noitemsep,topsep=0pt]
  \item replacing the total matter density $\rho_{\rm m}$ -- including neutrinos -- in the formula of the mass function of Eq.~\ref{eq:nm} by the matter density of cold dark matter and baryons only, $\rho_{\rm cdm+b}$. We will refer to this step alone as the ``matter prescription''.
  \item replacing the variance of the total matter density fluctuations as written in Eq.~\ref{eq:s2m} by the variance of dark matter $+$ baryons fluctuations only; this is equivalent to replacing the transfer function $T_{\rm m}(k,z)$ in the expression of the matter power spectrum in Eq.~\ref{eq:pk} by the  transfer function
  \begin{equation}
    T_{\rm cdm+b}= \frac{\Omega_{\rm cdm} T_{\rm cdm} + \Omega_{\rm b} T_{\rm b}}{\Omega_{\rm cdm}+\Omega_{\rm b}}
  .\end{equation}
\end{itemize}
The combination of these two steps will be referred to as the ``cold dark matter (CDM) prescription''. The two right panels of Fig.~\ref{fig:dndT} illustrate the influence of the two steps of the neutrino prescription on the cumulative temperature function. The effect of the first step (orange curve) is fairly straightforward, as it reduces the mass function of Eq.~\ref{eq:nm} by the ratio $\rho_{\rm cdm+b}/\rho_{\rm m}$, and thus the cumulative mass (or temperature) function by an amount that increases with the mass (or temperature). On the contrary, the second step (green curve) boosts the variance of matter fluctuations (described in Eq.~\ref{eq:s2m}) and therefore cluster abundances. When combined (red curve), the two steps somewhat balance each other, with a final net increase of cluster abundances. In the present work, we adopt this neutrino prescription whenever massive neutrinos are included in our analysis (using either one or both of the steps described).

Though not illustrated here, the impact of the phenomenological parameter $\gamma$ on the temperature distribution function is fairly straightforward. Values of $\gamma$ bigger that the fiducial $\Lambda$CDM one will decrease cluster abundances over the whole range of temperature or masses, as it lowers the whole matter power spectrum at once by scale-independent factor (cf. Eq.~\ref{eq:pkmod}). This distinguishes $\gamma$ from the effect of massive neutrinos, as the latter affect the matter power spectrum differently depending on the scales considered. As a consequence, although their global effect is the same, the two $\Lambda$CDM extensions considered here change cluster abundances in distinct ways, which will have consequences on our results (reported in Sect. \ref{sec:4-Res}).

On another practical note, the influence of a modification of gravity through $\gamma$ is implemented in our analysis once again through a modification of the matter power spectrum. Any value of $\gamma$ greater or smaller than $\sim0.545$ (corresponding to the best-fit value for $\Lambda$CDM) will result respectively in a smaller or greater growth rate $f$, which in turns impacts the growth factor $D$. By definition -- and neglecting potential scale dependencies -- the matter power spectrum is proportional to the square of the growth factor. Thus the resulting modified power spectrum can be written as
\begin{equation}\label{eq:pkmod}
  P_{\rm m,MG}(k,z,\gamma) = P_{\rm m}(k,z) \left(\frac{D(z_*)}{D(z)} \frac{D_{\rm MG}(z,\gamma)}{D_{\rm MG}(z_*,\gamma)} \right)^2
,\end{equation}
where $P_{\rm m}$ is the $\Lambda$CDM matter power spectrum, and $D$ and $D_{\rm MG}$ are respectively the growth factor in $\Lambda$CDM (i.e. for $\gamma\sim0.545$) and in the modified gravity scenario for the chosen $\gamma$, with both function normalized to 1 today. The redshift $z_*$ needs to be deep enough in the matter-dominated era, where $\Omega_{\rm m}\sim1$ and thus the influence of $\gamma$ is negligible; in practice, it is sufficient to choose $z_*\sim100$.

\section{Datasets and methods}\label{sec:3-DataMeth}


\subsection{Cluster and CMB data}\label{ssec:3.1-Data}

We use in our analysis the sample of X-ray selected clusters of \citet{2015A&A...582A..79I}, where a complete description of the catalogue can be found. We summarize its main characteristics here. This sample was built from the online database BAX \citep{2004A&A...424.1097S}. The redshift range of the sample was limited to $z=0.1$, so as to limit the potential internal evolution of the abundance of these objects. The chosen minimal X-ray flux ($=1.810^{-11}$ erg.s$^{-1}$.cm$^{-2}$) allows the sample to be complete. It is the largest ever used for the determination of the local temperature distribution function, with 73 clusters covering a temperature range of $[0.8,9]$ keV with a mean redshift of $z\sim0.05$.

The relation between the measured temperature of those clusters and their total masses is then given by the scaling law of Eq.~\ref{eq:scallaw} and a choice of calibration through the $A_{T-M}$ parameter. As mentioned earlier, $A_{T-M}$ is considered as a free parameter in most of our analysis.

However, in order to match the Planck mass calibration  \citet{2014A&A...571A..20P} (where the so-called hydrostatic bias value is $1-b=0.8$), we have to determine the corresponding value of $A_{T-M}$. By fitting X-ray-derived masses to the Planck SZ-derived ones, \citet{2015A&A...582A..79I} determined it to be approximately $\sim 7.86$, when using the same cluster mass definition (critical $M_{500}$) as \mbox{\citet{2014A&A...571A..20P}}.
In the following, we  refer to this specific value as the ``Planck cluster calibration''. To extrapolate this value to the main mass definition used in this work (the virial mass), we make use of the one-dimensional (1D) posterior distribution for $A_{T-M}$ that we produced for both mass definitions (cf. Fig.~\ref{fig:atm_like}). From it we can associate a ``probability'' (more precisely a marginalized likelihood value) to the Planck cluster calibration in the critical $M_{500}$ case. We then look for the $A_{T-M}$ value that has the same probability in the virial mass-case posterior, and is chosen as the ``Planck cluster calibration'' for this mass definition. We thus ensure that this resulting value of $A_{T-M}$ provides the same level of ``fitness'' (for the measured cluster temperature function) as in the critical $M_{500}$ case, while marginalizing over cosmological parameters. A consequence of this procedure is that the extrapolated $A_{T-M}$ value depends on the choice of the mass function. We derived as Planck cluster calibration values $A_{T-M} \sim 9.06$ and $A_{T-M} \sim 8.72$ when using the T08 and D16 mass functions, respectively. Although different, those two values are close enough to each other (well within the width of the 1$\sigma$ interval of the $A_{T-M}$ posterior) so that our choice will not affect much the relevant parts of our analysis. 

In combination with cluster data, we use the latest publicly available data release from the Planck Collaboration, namely the 2015 CMB spectra and associated likelihood code. We include both the TT, EE, BB, and TE likelihood in the low-multipole range and the TT, TE, and EE likelihood in the high-multipole range \citep[see][for a complete description]{2016A&A...594A..11P}.


\subsection{Additional probes}\label{ssec:3.2-AddProb}

To attempt to solve the observed tension between early and late times probes, we chose to consider two distinct $\Lambda$CDM extensions that both have a major effect at late times. While the Planck data provides exquisite constraints on the parameters of the standard cosmological model, CMB data alone tend to give relatively poor constraints on such extensions, as they mostly provide insight into the early Universe. As a consequence, we also test in the present work the robustness of our results when introducing two additional late-time probes on top of galaxy clusters:
\begin{itemize}[noitemsep,topsep=0pt]
  \item measurements of the BAO scale from the power spectrum of galaxies at high redshift from \citet{2014MNRAS.441...24A} and \citet{2014JCAP...05..027F};
  \item the 1D matter power spectrum reconstruction from \mbox{Lyman-$\alpha$} forest observations at an average redshift of $\sim 3.5$ from \citet{2015JCAP...11..011P}; for this probe, we adapted a code kindly made available to us by C.~Yèche and collaborators.
\end{itemize}
\mbox{Lyman-$\alpha$} data provides additional constraints on the shape of the power spectrum at different redshifts and valuable complementary information on structure formation, leading to tight cosmological constraints.


\subsection{Numerical methods and tools}\label{ssec:3.3-Meth}

To explore our full parameter space under the constraint of our datasets, we adopted a standard MCMC analysis using two publicly available codes: the CosmoMC package \citep{2002PhRvD..66j3511L, 2013PhRvD..87j3529L} and the Monte Python code \citep{Audren:2012wb}. In these codes, the computation of cosmological quantities and observables are performed respectively by the CAMB \citep{Lewis:1999bs,Howlett:2012mh} and CLASS \citep{2011JCAP...07..034B} Boltzmann codes, whose main purpose is to compute theoretical power spectra of CMB anisotropies.

The latest Planck CMB likelihood is already interfaced with the two aforementioned MCMC codes, as well as the BAO likelihood. The \mbox{Lyman-$\alpha$} likelihood was integrated by us into the MCMC codes, and adapted to the models considered in this work. We also developed a dedicated module for computing the likelihood associated with our cluster sample; the module makes use of the outputs from the Boltzmann codes (such as background quantities and matter power spectrum) to compute the required cluster observables. A more detailed description of technical aspects of this likelihood module can be found in \citet{2015A&A...582A..79I}.

While massive neutrinos are already implemented in both the CLASS and CAMB codes, the other $\Lambda$CDM extension we considered (the $\gamma$ parametrization) is not part of these Boltzmann codes by default. As a model introduced to study phenomenological modifications of the late growth of structure, it is not very well suited for CMB studied and is tricky to implement consistently in a Boltzmann code, compared to a proper full theory of modified gravity. In the present work we therefore implement the $\gamma$ model only at the level of the likelihood for two of our affected probes, namely clusters and \mbox{Lyman-$\alpha$} observations, by modifying the matter power spectrum according to Eq.~\ref{eq:pkmod}. The amplitude of matter fluctuations is therefore given by
\begin{equation}\label{eq:sigmodgam}
    \sigma_{M,MG}(z)=\frac{D(z_*)}{D_{MG}(z_*)}\frac{D_{MG}(z)}{D(z)}\,\sigma_{M,\Lambda}(z)
\end{equation}
and we then compute the mass function according to our prescriptions as explained above. We leave out from the main analysis the effects of a late modification of gravity on the CMB observables, which would appear in the power spectra of its anisotropies mainly through lensing and ISW effects. We explore briefly in Appendix~\ref{sec:Appendix-B} if accounting for one of these effects (ISW) can affect our conclusions, although a more thorough treatment would require going beyond the $\gamma$ parametrization as it is not well-suited for CMB studies. Finally we chose to neglect correlations between our probes, arguing that they should be fairly low as our datasets do not overlap much in redshift. Consequently, the total combined likelihood of all our probes will be simply the product of all the individual likelihoods considered.

On a more technical note, throughout our work we use flat priors for the six standard $\Lambda$CDM cosmological parameters (the same ones used in the Planck CMB analyses), as well as for the sum of neutrino masses, our calibration parameter $A_{T-M}$ , and our MG parameter $\gamma$. In most of the scenarios we explored, we checked a posteriori that our choice of priors did not influence our posterior distributions. In the cases where they did have an influence, we mention it in the relevant section.


\section{Results}\label{sec:4-Res}


\subsection{(Re-)stating the tension in $\Lambda$CDM}\label{ssec:4.1-Tens}

\begin{figure}[t]
    \centering
    \includegraphics[width=\columnwidth]{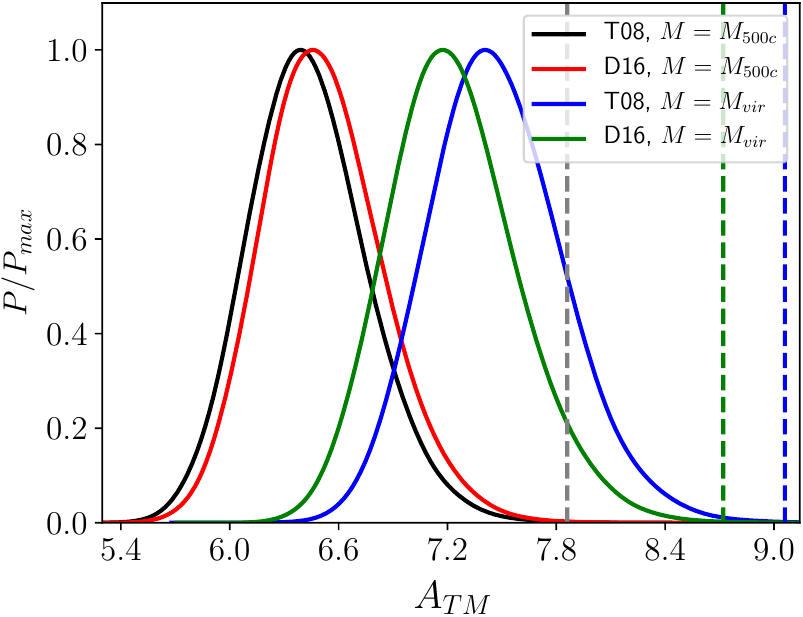}
    \caption{Posterior distributions for the mass calibration parameter $A_{T-M}$ when combining CMB and X-ray cluster data, for the T08 and D16 mass function and two definitions for the cluster mass ($M_{500c}$ and $M_{vir}$). The grey vertical dashed line corresponds to the Planck cluster calibration value for the $M_{500c}$ mass definition. The blue and green vertical dashed lines correspond to the extrapolated Planck cluster calibration for the $M_{vir}$ definition, using respectively the T08 and D16 mass function.}
    \label{fig:atm_like}
\end{figure}

\begin{figure}[t]
    \centering
    \includegraphics[width=\columnwidth]{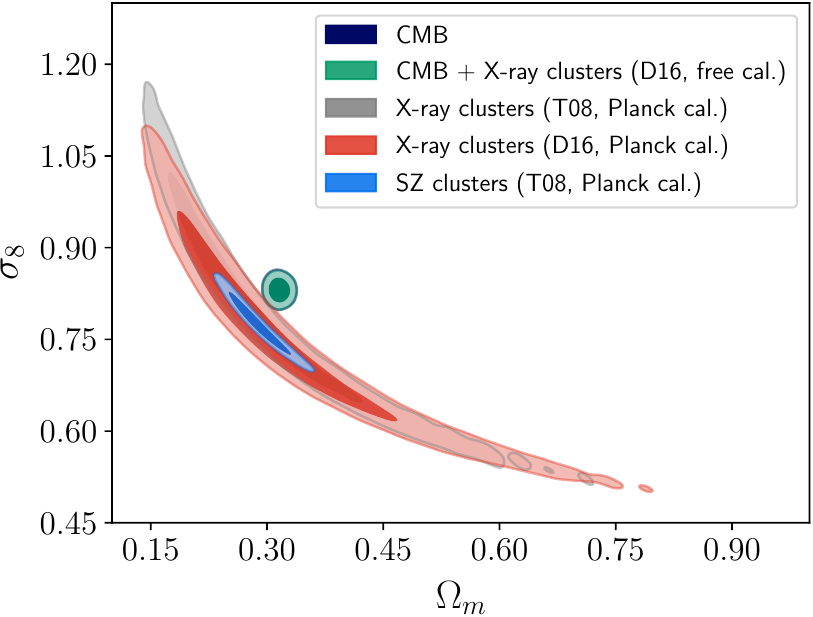}
    \caption{Confidence contours (68/95\%) in the $\Omega_m-\sigma_8$ plane for various combinations of datasets (CMB, X-ray or SZ clusters) and assumptions (T08 or D16 mass function, $A_{T-M}$ left free or fixed to the Planck cluster calibration, cf. Sect.~\ref{ssec:3.1-Data}). For the clusters-only cases, additional uniform priors on $\Omega_m$ and $H_0$ have been applied (see Sect.~\ref{ssec:4.1-Tens} for details). On this figure, the ``CMB'' and ``CMB + X-ray clusters'' contours are virtually identical.}
    \label{fig:om_vs_s8} 
\end{figure}

The tension between cluster counts and the Planck CMB measurements in the $\Lambda$CDM paradigm is best illustrated by their respective constraints in the $\Omega_m - \sigma_8$ plane, as shown in Fig.~\ref{fig:om_vs_s8}. In this plane, the CMB alone (green contour) provides tight constraints that represent a tiny part of the parameter space. On the other hand, X-ray cluster counts are not sensitive to the Hubble constant and are almost independent of the other cosmological parameters except for the amplitude of matter fluctuations at a fixed matter density. This leads to a band of degeneracy (red and blue contours respectively for the T08 and D16 mass functions) between the two parameters. When using only clusters as constraints, we imposed additional flat priors on both $\Omega_m$ ($\in [0,1]$) and $H_0$ ($\in [20,100]$) to help the convergence of our MCMC. Those priors do influence the posterior contours of Fig.~\ref{fig:om_vs_s8} (and Fig.~\ref{fig:om_vs_s8_mnu} in the next section), with wider priors resulting in more extended contours but only in the direction of the $\Omega_m - \sigma_8$ degeneracy. Therefore, these additional priors do not affect our conclusions: using here the standard Planck mass calibration for our X-ray clusters, we clearly see a significant tension between the constraints from the two datasets, entirely consistent with the one obtained from Planck SZ counts \citep{2014A&A...571A..20P} and the conclusion of \citet{2015A&A...582A..79I}. This conclusion is virtually independent of the choice of mass function (T08 or D16). Relaxing the constraint on $A_{T-M}$, the tension can then be directly visualized with the 1D posterior distribution of $A_{T-M}$ shown in Fig.~\ref{fig:atm_like}: as expected, the standard Planck cluster calibration (dashed line) is quite strongly disfavoured by the data (at the $\sim 4 \sigma$ level for the T08 mass function and critical $M_{500}$ mass definition). This leads again to the conclusion that either our cosmological model or our understanding of clusters needs to be revised in order to reconcile both datasets.

\subsection{Introducing massive neutrinos}\label{ssec:4.2-Resnu}


\subsubsection{X-ray clusters versus CMB}\label{ssec:4.2.1-}

We present our results when introducing three degenerate, massive neutrinos in the $\Lambda$CDM model, with the sum of their mass $\sum m_\nu$ as an additional free parameter (as opposed to a single massive one with mass 0.06 eV and two massless in the fiducial $\Lambda$CDM model). 

\begin{figure}[t]
  \centering
  \includegraphics[width=\columnwidth]{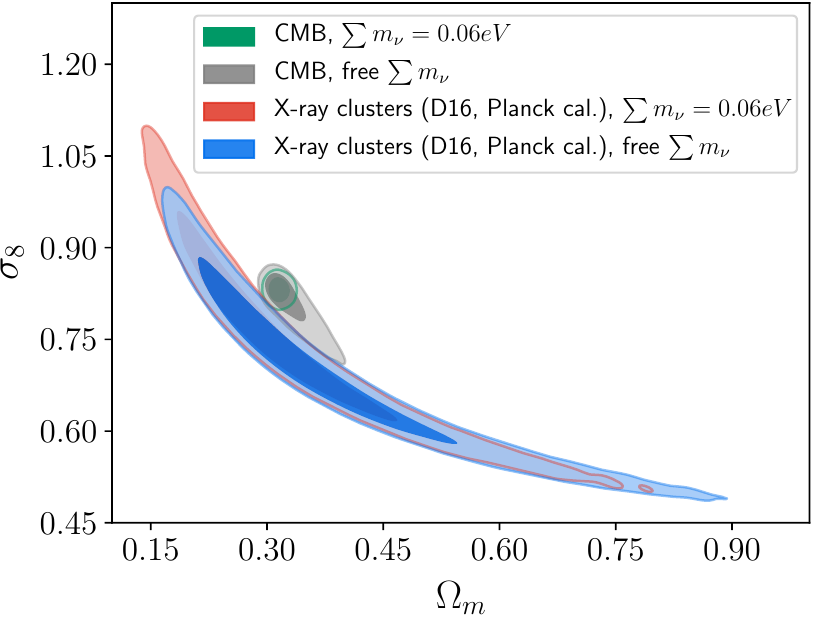}
  \caption{Confidence contours (68/95\%) in the $\Omega_m - \sigma_8$ plane for various datasets (CMB and X-ray clusters) and assumptions (three neutrinos with free masses or only one fixed to the standard 0.06 eV value). For the clusters-only cases, $A_{T-M}$ is fixed to the Planck cluster calibration and additional uniform priors on $\Omega_m$ and $H_0$ have been applied (see Sect.~\ref{ssec:4.1-Tens} for details). The ``CMB + X-ray clusters'' contours with free calibration (not shown on this figure) are again identical to the ``CMB'' contours.}
  \label{fig:om_vs_s8_mnu}
\end{figure}

We initially kept the normalization parameter $A_{T-M}$ fixed to its standard Planck  cluster value.
Results are summarized in the $\sigma_8-\Omega_m$ plane of Fig.~\ref{fig:om_vs_s8_mnu}, which shows on the one hand CMB-alone constraints, and on the other hand the contours produced by \mbox{X-ray} clusters assuming the standard Planck cluster calibration, both with and without free neutrino masses. In the case of clusters, the effect of inclusion of free neutrino masses is to shift slightly the crescent-shaped contours towards lower $\sigma_8$ and higher $\Omega_m$ along the main degeneracy direction, not modifying the preferred value of $\sigma_8$ for a fixed $\Omega_m$. This effect is quite small and comparable to the difference we found when using the two different mass function (T08 and D16). For the CMB, allowing for massive neutrinos opens up slightly the contours (from green to grey) towards lower amplitudes of the matter fluctuations $\sigma_8$ and higher $\Omega_m$. Because the CMB contours are essentially parallel to those of clusters, the shift produced by massive neutrinos on both contours does not help to reduce the tension. We conclude that the tension between CMB and X-ray cluster counts with the standard Planck mass calibration cannot be alleviated by allowing for massive neutrinos: in the $\Omega_m \sim 0.3$ region, the inclusion of massive neutrinos leaves the cluster contours essentially unchanged, while the CMB contours open up in a way parallel to the cluster contours.

\begin{figure}[t]
    \centering
    \includegraphics[height=0.5\columnwidth]{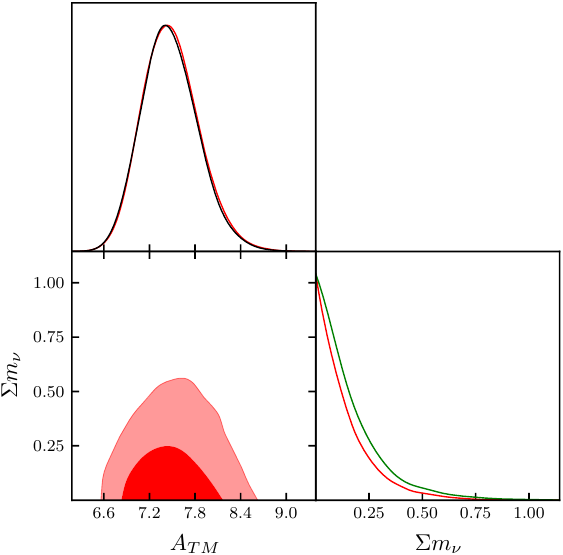}\includegraphics[height=0.5\columnwidth]{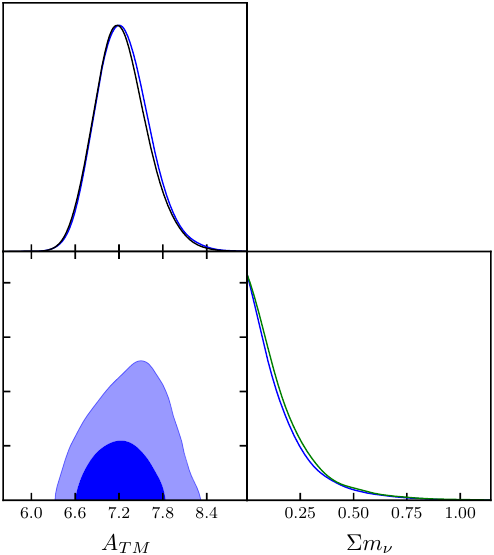}
    \caption{Confidence contours (68/95\%) and posterior distributions for the $A_{T-M}$ and $\sum m_{\nu}$ parameters for our two choices of mass functions (T08 in red and D16 in blue) with the $M_{vir}$ cluster mass definition. The green $A_{T-M}$ posteriors correspond to the case where the neutrino mass is fixed to the standard 0.06 eV value. The black $\sum m_{\nu}$ posterior comes from the CMB-only case.}
    \label{fig:plkbaxATM3nu}
\end{figure}

\begin{figure}[t]
    \centering
    \includegraphics[width=\columnwidth]{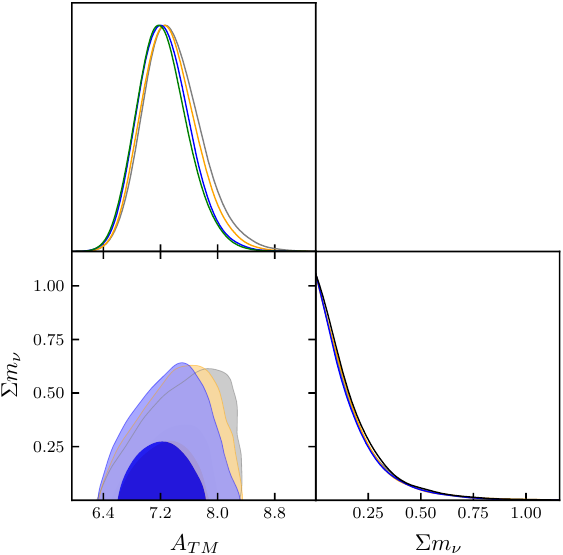} 
    \caption{Confidence contours (68/95\%) and posterior distributions for the $A_{T-M}$ and $\sum m_{\nu}$ parameters, using the D16 mass function and $M_{vir}$ cluster mass definition. Three choices of neutrino prescription are shown: CDM prescription (blue), no prescription (yellow), and matter prescription (grey). The same conventions for the green and black posteriors are used as in Fig.~\ref{fig:plkbaxATM3nu}.}
    \label{fig:plkbaxATM3nupres}
\end{figure}

Let us now examine in more detail this issue when we relax the mass calibration. In Fig.~\ref{fig:plkbaxATM3nu} we present the contours in the $A_{T-M}-\sum m_{\nu}$ plane for the T08 (red) and D16 (blue) mass functions. The 1D posteriors on the neutrino masses as well as on $A_{T-M}$ are almost unchanged. We also examine the role of the various prescriptions in Fig.~\ref{fig:plkbaxATM3nupres}: small differences are found that remain well below the $1 \sigma$ uncertainty. The contours show a weak correlation of $A_{T-M}$ with neutrino masses: going from a mass of $\sim$ 0 to 0.5 eV, the preferred value of $A_{T-M}$ is increased by $\sim 0.2$, $0.3$ and $0.7$ respectively in the CDM prescription, no prescription, and matter prescription cases. These small effects seem to vary slightly with the mass function and are far too small (well below the $\sim$ 1.5 width of the 68\% confidence limits) to modify the amplitude of the discrepancy: the likelihoods on $A_{T-M}$ and on the neutrino masses are essentially unchanged. In the following, we  adopt the CDM prescription as a reference.

When massive neutrinos are allowed, contours in the $\sigma_8-\Omega_m$ plane are almost identical in the range of $\Omega_m$ preferred by the CMB. We conclude that there is no indication that neutrinos being massive would allow any appreciable release of the so-called clusters-CMB tension.

\subsubsection{Combining with additional probes}\label{ssec:4.2.2}

Stringent constraints on cosmology are obtained through the combination of different probes. Here we briefly examine the effects of adding two probes that directly constrain the power spectrum shape presented in Sect.~\ref{ssec:2.4-ClusObs}. We followed the same procedure as in the previous section, now constraining the evolution of our MCMC with clusters, CMB, BAO, and \mbox{Lyman-$\alpha$} data. We allow for massive neutrinos and a free calibration $A_{T-M}$ in addition to the usual $\Lambda$CDM cosmological parameters. As shown before, given that our results are barely sensitive to the choice of the mass function, we  work only with the more recent D16 mass function in the following. 

\begin{figure}[t]
    \centering
    \includegraphics[width=0.49\columnwidth]{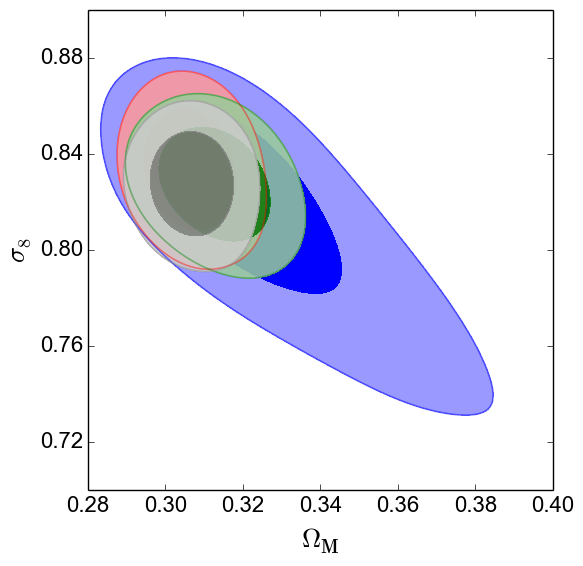}
    \includegraphics[width=0.49\columnwidth]{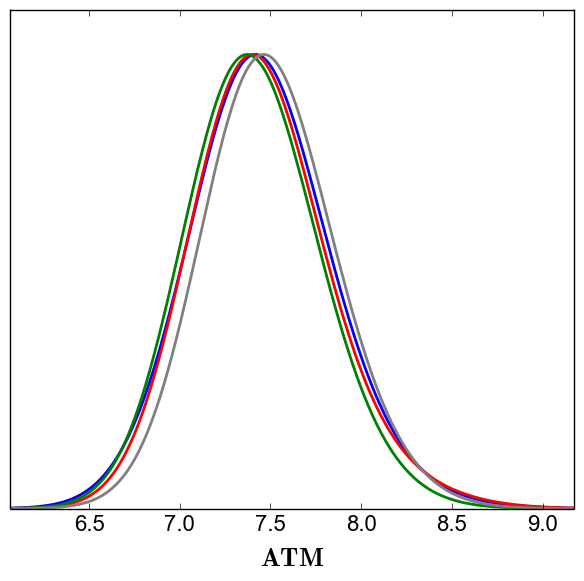}
    \includegraphics[width=0.49\columnwidth]{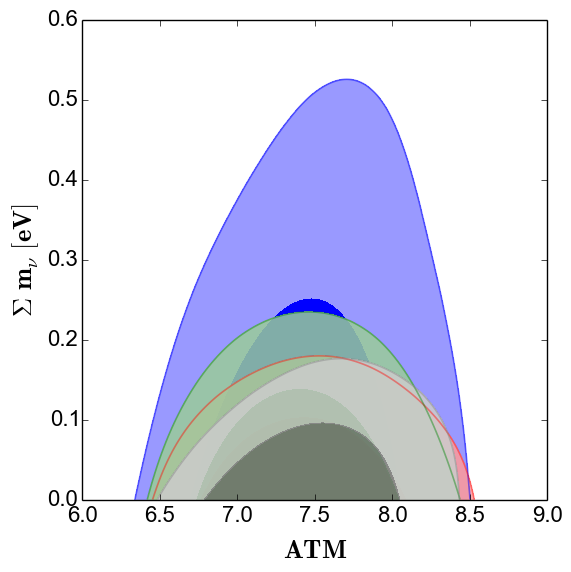}
    \includegraphics[width=0.49\columnwidth]{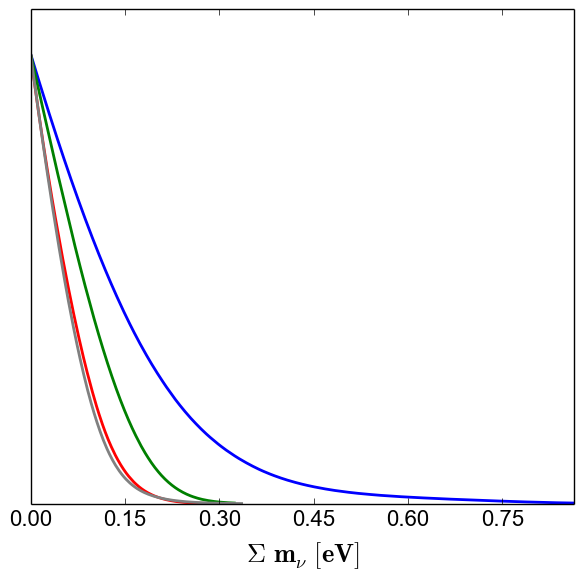}
    \caption{Confidence contours (68/95\%) and posterior distributions for the $\Omega_m$, $\sigma_8$, $A_{T-M}$ , and $\sum m_{\nu}$ parameters, using the D16 mass function and $M_{vir}$ cluster mass definition. We show results obtained from CMB and clusters data (blue) and the effects of adding \mbox{Lyman-$\alpha$} data (green), BAO data (red), and both (grey).}
    \label{fig:plklyabaxATMnu} 
\end{figure}

Our results are illustrated in Fig.~\ref{fig:plklyabaxATMnu}. While clusters with a free calibration do not change noticeably constraints on neutrinos masses (as already seen in Fig.~\ref{fig:plkbaxATM3nu}), the addition of BAO and \mbox{Lyman-$\alpha$} data leads to tighter constraints in agreement with \citet{2015JCAP...11..011P} and \citet{2017JCAP...06..047Y}. On the other hand, the addition of these datasets essentially leads to $A_{T-M}$ constraints identical to the CMB alone.

\subsection{Modified gravity as an alternative solution}\label{ssec:4.3-Resgam}

Our detailed investigation in the previous section led to the firm conclusion that the tension between CMB and clusters with the Planck cluster calibration remains unsolved when neutrino masses are left free. Of course the simplest solution is to consider that the cluster mass calibration is to be revised downwards to a value $A_{T-M} \sim 7.2$ corresponding to $1-b \sim 0.6$. This, however, leads to high masses for clusters that are above most observational estimations and to a high amplitude of matter fluctuations, thus alternative possibilities are to be considered. In this section we  examine a second possibility, namely that the late growth of structures does not follow the predictions of the standard $\Lambda$CDM model, but results from modified laws of gravity. To do so, we consider a phenomenological modification of the late growth of structure controlled by the introduction of a new free parameter $\gamma$ as described in Sect.~\ref{ssec:2.3-MG}. As mentioned in Sect.~\ref{ssec:2.4-ClusObs}, we consider here only the implication of the modification of the growth rate on the cluster mass function and not the CMB observables. We follow the same approach as in previous sections: a standard MCMC analysis to explore our full parameter space under the constraints of our datasets, with the standard cosmological parameters and the calibration $A_{T-M}$ being free and the index $\gamma$ as an additional parameter of the model.


\subsubsection{CMB and clusters constraints}

We can already intuit the constraints resulting from the combination of X-ray clusters data and the CMB data in this new paradigm. Given that the calibration $A_{T-M}$ is left free, the amplitude of matter fluctuations is essentially unconstrained by cluster abundance and a degeneracy between $A_{T-M}$ and $\gamma$ can be expected. We also expect constraints on other cosmological parameters to remain essentially unchanged. We extended our analysis by introducing again non-zero masses for neutrinos: we checked in various cases that all constraints on cosmological parameters are unchanged except those on $\sigma_8$ and thereby on $A_{T-M}$ and $\gamma$. The parameters $A_{T-M}$ and $\gamma$ are highly degenerated in the range of interest ($ 0 < \gamma < 1$).  We found that the (degenerated) contours in the $A_{T-M} - \gamma$ plane remain remarkably stable. Figure~\ref{fig:plkbaxATMgamwo3nu} summarizes the relation between those two parameters in a variety of models explored in this work and with different datasets. The choice of flat prior on $\gamma$ ($\in [0, 1]$ in our case) has an influence on our posterior distributions: it limits the extent of the more degenerated contours of Fig.~\ref{fig:plkbaxATMgamwo3nu}, but it does not impact our discussions.


\subsubsection{Adding BAO and \mbox{Lyman-$\alpha$} data}

\begin{figure}[t]
    \centering
    \includegraphics[width=0.6\columnwidth]{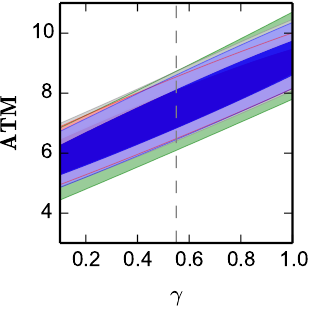}
    \caption{Confidence contours (68/95\%) in the $A_{T-M} - \gamma$ plane from CMB and clusters data, with (grey) or without (green) free neutrino masses. The D16 mass function and $M_{vir}$ cluster mass definition are used here. The $A_{T-M} - \gamma$ correlation remains stable when adding BAO (red) or \mbox{Lyman-$\alpha$} (blue) data (neutrinos masses still free).}
    \label{fig:plkbaxATMgamwo3nu}
\end{figure}

As seen in previous sections, the addition of these two probes mainly limits the range of allowed neutrino masses, but has a limited impact on the calibration $A_{T-M}$. Consequently, the degeneracy between $A_{T-M}$ and $\gamma$ is expected to remain unchanged. This is indeed verified in Fig.~\ref{fig:plkbaxATMgamwo3nu} where the contours in the $A_{T-M} - \gamma$ plane are almost identical.

\begin{figure}[t]
    \centering
    \includegraphics[width=\columnwidth]{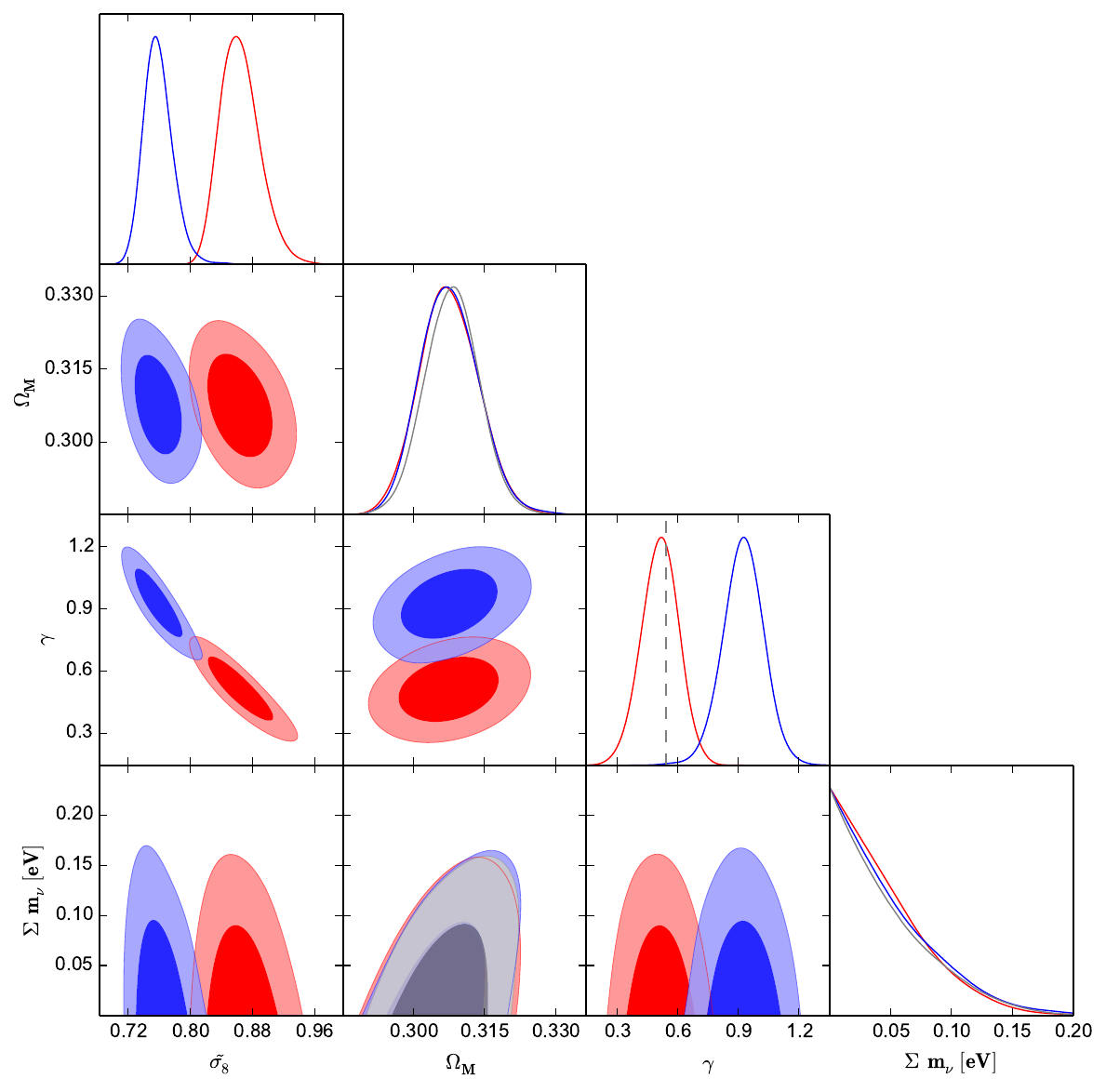}
    \caption{Confidence contours (68/95\%) and posterior distributions for the $\Omega_m$, $\sigma_8$, $\gamma,$ and $\sum m_{\nu}$ parameters, using the D16 mass function and $M_{vir}$ cluster mass definition. Results are obtained from the combination of CMB, clusters, BAO, and \mbox{Lyman-$\alpha$} data. We fix $A_{T-M}$ to either the Planck cluster calibration value (blue) or its preferred value when left free (red, see Sect.~\ref{ssec:4.4} for details). We also show in grey the posteriors in the standard gravity case with no cluster data. The grey vertical dashed line corresponds to the standard $\gamma$ value of $\sim 0.545$.}
    \label{fig:plkbaxbaoLAgam3nutwoATM}
\end{figure}


\subsection{Concluding on the calibration issue}\label{ssec:4.4}

As a final illustration of our analysis, we ran a MCMC combining CMB, clusters, BAO and \mbox{Lyman-$\alpha$} data, with both neutrino masses and the growth rate index $\gamma$ being free and with two fixed values of the calibration $A_{T-M}$:
\begin{itemize}[noitemsep,topsep=0pt]
  \item $A_{T-M}~=~7.19$, the preferred value for $\Lambda$CDM from our joint analysis of CMB and clusters using the D16 mass function;
  \item $A_{T-M}~=~8.72$, the value corresponding to the Planck cluster calibration.
\end{itemize}
The results of this comparison are shown in Fig.~\ref{fig:plkbaxbaoLAgam3nutwoATM} where the red and blue contours correspond respectively to the first and second choice for the value of $A_{T-M}$. In addition, we plotted the 1D posterior distribution of $\Omega_m$ and $\sum m_{\nu}$ in the standard-gravity case. Both the $\sum m_{\nu} - \Omega_m$ contours and the corresponding 1D posteriors are virtually identical. On the other hand, the posteriors on $\gamma$ and $\sigma_8$ show strong differences: the Planck cluster calibration leads to a high preferred value of $\gamma\sim0.9$ inconsistent with the standard model, while the other calibration choice leads to $\gamma=0.55\pm0.08$ in complete agreement with $\Lambda$CDM expectations.


\section{Conclusion}\label{sec:Conclusions}

In the present paper, we examined the discrepancy on the amplitude of matter fluctuations as estimated by $\sigma_8$ obtained from X-ray cluster abundance on one side, and derived from the Planck CMB fluctuations in $\Lambda$CDM on the other side. Two possible extensions of the standard $\Lambda$CDM were examined: the presence of massive neutrinos and the impact of a modification of gravity on the growth rate. Our strategy was to examine the constraints that CMB and cluster abundance data yield, without further additional assumptions or data on clusters, that is, leaving the calibration of the mass temperature relation $A_{T-M}$ free. Using the sole combination of X-ray clusters and Planck CMB data, we found no appreciable correlation between the cluster mass calibration and neutrinos masses (with $\sum m_\nu \lesssim 0.47$ eV at the 95\% confidence level).

The addition of BAO constraints as well as those provided by the 1D \mbox{Lyman-$\alpha$} forest spectrum allows tighter constraints to be imposed on the sum of the neutrino masses while leaving the calibration essentially unchanged compared to the massless case. From this we firmly conclude that the neutrino masses do not relax the CMB-clusters tension in the standard cold dark matter picture.
Indeed, when we compare constraints obtained with two different calibrations -- the standard Planck cluster calibration and a calibration based on matching cluster abundance in a CMB normalized $\Lambda$CDM model -- we found essentially the same constraints on neutrinos masses. This also leads to the conclusion that the CMB-cluster tension is closely related to the cluster mass calibration issue. Recently, \citet{2018A&A...614A..13S} have revised the constrained on the $1-b$ calibration parameter using Planck SZ cluster counts and the power spectrum of the hot gas, in addition to the Planck CMB data using a revised value of the optical depth $\tau$ from \citet{2016A&A...596A.108P}. Their conclusions on the calibration and its role in solving the tension  are in qualitative agreement with our conclusions.

Introducing the $\gamma$ model for the growth rate as a simple modification of gravity, we found a tight correlation between $A_{T-M}$ and $\gamma$. This correlation appeared to be insensitive to the presence of possible massive neutrinos or the addition of complementary data. We conclude that the CMB-cluster tension cannot be solved simply by non-zero masses for neutrinos: if the Planck cluster mass calibration is to be consolidated, this would be a strong indication that the simple model of $\Lambda$ cold dark matter with standard massive neutrinos cannot accommodate present data and would call for new physics in the dark sector.

\begin{acknowledgements}

ZS was supported by a grant of excellence from the Agence des Universités Francophones (AFU). SI was supported by the European Structural and Investment Fund and the Czech Ministry of Education, Youth, and Sports (Project CoGraDS - CZ.02.1.01/0.0/0.0/15\_003/0000437). This work has been carried out thanks to the support of the OCEVU Labex (ANR-11-LABX-0060) and the A*MIDEX project (ANR-11-IDEX-0001-02) funded by the ``Investissements d'Avenir'' French government programme managed by the ANR. It was also conducted using resources from the IN2P3 Lyon computing center. We acknowledge C.~Yèche and collaborators for providing us with their code to implement \mbox{Lyman-$\alpha$} constraints. 

\end{acknowledgements}

\bibliographystyle{aa}
\bibliography{Bibliography}


\appendix

\section{Effects of additional degrees of freedom in the neutrino sector}\label{sec:Appendix-A}

\begin{figure}[t]
  \centering
  \includegraphics[height=0.49\columnwidth]{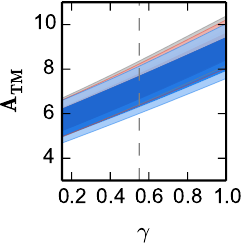}
  \includegraphics[height=0.49\columnwidth]{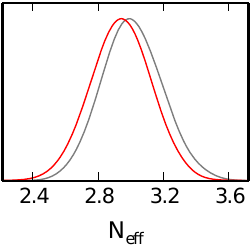}  
  \caption{Confidence contours (68/95\%) and posterior distributions for the $A_{T-M}$, $\gamma$ , and $N_{\rm eff}$ parameters, using the D16 mass function and $M_{vir}$ cluster mass definition. Results are obtained from the combination of CMB and cluster data when the $N_{\rm eff}$ parameter is either fixed to 3.046 (blue) or left free (red). The effects of adding BAO data in the latter case are shown in grey.}
  \label{fig:plkbaxgamatmnucdmexoneff} 
\end{figure}

In this appendix, we consider extra relativistic relics by treating the effective number of neutrinos $N_{\rm eff}$ as a new parameter, and/or by introducing a massive sterile neutrino into the model (with a free parameter $m_s$ controlling its mass). We first considered a model where both the $\gamma$ and the calibration $A_{T-M}$ are left free, as well as the $N_{\rm eff}$ parameter. Figure~\ref{fig:plkbaxgamatmnucdmexoneff} shows our results when using CMB and cluster data with (grey) and without (red) the addition of BAO data. We find constraints consistent with our previous results where $N_{\rm eff}$ was fixed to its 3.046 fiducial value (blue). The $\gamma - A_{T-M}$ correlation remains, while $N_{\rm eff}$ appears to vary independently of these two parameters. Although slightly tightening the constraints, adding BAO data does not change those results.

\begin{figure}[t]
  \centering
  \includegraphics[width=0.55\columnwidth]{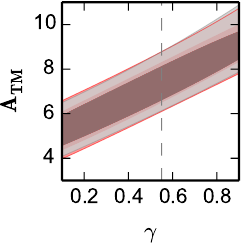}
  \caption{Confidence contours (68/95\%) in the $A_{T-M} - \gamma$ plane, using the D16 mass function and $M_{vir}$ cluster mass definition. Results are obtained from the combination of CMB and cluster data when the $N_{\rm eff}$ parameter is either fixed to 3.046 with massive active neutrinos (red) or left free to vary adding massive active neutrinos and a sterile neutrino with free mass (grey).}
  \label{fig:plkbaxgamatmnucdmexo}
\end{figure}

We finally considered a model where $N_{\rm eff}$ could change as a result of the presence of a sterile neutrino. Mentioned in Sect.~\ref{ssec:2.2-nu}, such a type of neutrino adds another degree of freedom to our problem, and could thus increase the chances of fixing the CMB-clusters tension. We followed an approach similar to \citet{2016A&A...594A..13P} to include both active massive neutrinos and a sterile one. The corresponding results shown in Fig.~\ref{fig:plkbaxgamatmnucdmexo} illustrate that the  $\gamma - A_{T-M}$ correlation remains untouched.

\section{Including a modification of the ISW effect} 
\label{sec:Appendix-B}

As mentioned in Sect.~\ref{ssec:3.3-Meth}, any change in the evolution of the growth rate of structures (e.g. through a modification of gravity or additional ingredients in our cosmological model) affects the angular power spectra of the CMB, mainly through the linear ISW effect at large scales and the gravitational lensing on smaller scales. While these effects have been properly accounted for when we explored the addition of massive neutrinos in our model, we did not include the effects of our phenomenological modification of gravity on the CMB power spectra. Indeed, the $\gamma$ model has been designed for late probes of the growth of structures and is much less appropriate for CMB and early universe predictions.

\begin{figure}[t]
  \centering
  \includegraphics[width=0.3\textwidth]{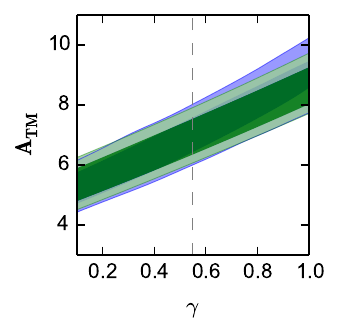}
  \caption{Confidence contours (68/95\%) in the $A_{T-M} - \gamma$ plane, using the D16 mass function and $M_{vir}$ cluster mass definition. Results are obtained from the combination of CMB and clusters data when accounting (green) or not (blue) for the modification of the ISW effect in the CMB power spectra due to our phenomenological modification of gravity.}
  \label{fig:plkbaxgamtildeatmisw}
\end{figure}

In order to assess whether neglecting those effects could affect our conclusions, we applied a recipe to include nonetheless one of them -- the ISW effect, much simpler to account for -- and modified the Boltzmann codes we used accordingly. We show our results in Fig.~\ref{fig:plkbaxgamtildeatmisw}: we observe no significant effect on the $A_{T-M} - \gamma$ correlation nor on their respective posterior distributions (not shown). Our conclusions thus remain unchanged.

\end{document}